\documentclass{elsart}

\usepackage{graphicx}

\usepackage{amssymb}

\def\GdBaPrd{$\rm{Gd(Ba_{2-x}Pr_x)Cu_3O_{7-\delta}}$ }

\def\GdPrd{$\rm{Gd_{1-x}Pr_xBa_2Cu_3O_{7-\delta}}$ }

\def\RPrd{$\rm{R_{1-x}Pr_xBa_2Cu_3O_{7-\delta}}$ }
\def\R{$\rm{RBa_2Cu_3O_{7-\delta}}$ }

\def\CuO{$\rm{CuO_2}$ }

\begin{document}

\begin{frontmatter}



\title{ $T_C$ suppression in Pr-doped 123-systems: Hole clustering }


\author[iust,alzahra]{V. Daadmehr\corauthref{cor1}},
\author[iust]{M.D. Niry} and
\corauth[cor1]{Corresponding author.} \ead{daadmehr@iust.ac.ir}
\author[sharif2]{A.T. Rezakhani}
\address[iust]{Magnetic and Superconductivity Research Lab, Department of Physics, Iran University of Science and Technology, P.O. Box 16844, Tehran, Iran}
\address[alzahra]{Magnetic and Superconductivity Research Lab, Department of Physics, Al.Zahra University, Tehran, Iran}
\address[sharif2]{Department of Physics, Sharif University of Technology, P.O. Box 11365-9161, Tehran, Iran}

\begin{abstract}
Doping Pr in 123-systems gives rise to some anomalies, such as $T_C$
suppression. Here, we show that a modification of hole localization
theory based on a geometrical modeling by band percolation theory
can put forward a good explanations for $T_C$ suppression for whole
range of Pr-doping value. In this model the central effect of hole
clustering is introduced. Also we have provided some experimental
evidences which manifest an agreement between simulation based on
hole clustering effect and the experimental data.
\end{abstract}


\end{frontmatter}

\section{Introduction}\label{secIntroduction} Pr substitutions and its effects on
suppression of superconductivity, pinning effect, critical current
density $(J_C)$, and other transportation properties in \GdPrd
(GdPr-123) compounds are among the most interesting subjects in
High-Temperature Superconductors (HTSC). The series of \R compounds,
where R is Y or another rare-earth element except cerium (Ce),
praseodymium (Pr), promethium (Pm) and terbium (Tb) are metallic and
superconducting, with critical temperatures $(T_C)$ that range from
90 to 94 $K$ \cite{Akhavan3}. These compounds commonly have an
orthorhombic layered perovskite-like structure, containing \CuO
planes within which there are mobile holes that are believed to play
role in superconductivity \cite{TTHN}. However the Pr-123 systems
has orthorhombic phase structure but it is not a superconductor
\cite{Akhavan3}. Some theoretical models such as, hole filling
\cite{Matsuda,Gonacalves}, pair breaking \cite{Kebede,Neumeier}, and
localized holes \cite{Tang,Fehrenbacher,Stutzman} have been
presented to explain the suppression of $T_C$ and
superconducting-insulator transition in these compounds. Recently,
some groups have reported observation of superconductivity in Pr-123
systems \cite{Mohammadizadeh3}. Blackstead et al. \cite{Blackstead}
have claimed that mis-substitution of Pr in RPr-123 causes the
suppression of superconductivity. In our previous work
\cite{Daadmehr}, it has been shown that in spite of the substitution
of Pr ion in the rare-earth site, superconductivity is suppressed in
GdPr-123 compounds. Researchers have tried to explain this effect by
some mechanisms such as hole filling, hybridization,
mis-substitution, hole localization \cite{Radosky}. Although, the
hole filling, hybridization of Pr $4f$ and O $2p$, and
mis-substitution cannot explain completely $T_C$ suppression effect
\cite{Radosky}, the hole localization is a good base for theoretical
explanation of this effect \cite{Tang,Fehrenbacher,Muroi2}. In
addition, we note that the hole localization have a good agreement
with experimental data for lower Pr-doping values \cite{Muroi2}.\\
The hole concentration on \CuO planes decreases by increasing
Pr-doping value in these compounds \cite{Akhavan3,Muroi2}. As well,
oxygen deficiency effect in these samples results in decrease in
hole concentration \cite{Daadmehr}. Indeed, it has been shown that
there is a correlation between Pr-doping value (x) and oxygen
deficiency $(\delta)$ \cite{Akhavan}. Moreover, there are some
dopants such as Ca and Sr which give rise to an increase of hole
concentration in \CuO planes, which instead results to an increase
in $T_C$. Obviously, the localization model can justify the
suppression of $T_C$ in lower values of Pr-doping and it cannot
explain other ranges of Pr-doping values, and then need to be corrected.\\
In this paper, we investigate the behavior of hole concentration in
\CuO planes in 123-systems for higher Pr-doping values wherein phase
transition occurs. This phenomena subject can be related to Coulomb
repulsion of localized holes, the collective behavior of Pr ions and
clustering effects of Pr-doping on \CuO planes in these systems.
Here, we try to find a good agreement experimental data and hole
localization and, also, hole collective effect of Pr in 123-system
for the range of $x$ from 0 to $x_C$, where superconductivity
completely suppress. Beyond this critical value of doping, the
system experiences a transition to the normal sate. in this case,
however, full analysis of the system requires a 3D percolation
modeling.\\
Our investigation, in this paper, is based upon geometrical
considerations. By using concept of band percolation we have
introduced notion of hole clustering and hole trapping by which we
have modified the effects of localization. In this way, it has been
shown that one obtains a good agreement between experimental data
and hole collective effects, and also these effects can be extended
to full range of Pr-doping.

\section{Experimental Details}\label{secExperimental}Single-phase polycrystalline \GdPrd samples, with $0<x<1$
stoichiometry have been prepared by standard solid-state reaction
technique. The mixture of $\rm Gd_2O_3$, $\rm Pr_6O_{11}$, $\rm
BaCO_3$, and $\rm CuO$ powder with $99.9 \%$ purity have been
calcined at $840\,^\circ C$ in air atmosphere for $24~h$. The
calcination were repeated twice with intermediate grinding. The
compounds were then pressed into pellets and sintered in oxygen
atmosphere at $930\,^\circ C$ for $36~h$. To get a sample containing
predominately orthorhombic structure we oxygenated it in
$650\,^\circ C$ for a $1~h$ period, so that $(\delta \thickapprox
0.15)$. Indeed,
the preparation procedure is similar to the one of our previous reports \cite{Daadmehr}.\\
The sintered samples were characterized by powder X-ray diffraction
(XRD) using $CoK\alpha$ radiation and scanning electron microscopy
(SEM) experiments. The resistivity measurements were carried out by
standard ac four-probe method, with 10 mA ac current and frequency
of 33 Hz using PAR-124 Lock-in amplifier. The temperature
measurements were made by a Lake Shore-330 temperature controller
with a Pt resistor and a GaAs diode. The temperature in the range of
$300-10~K$ with $0.10~K$ accuracy was provided by helium
closed-cycle refrigerator.

\section{Results and Discussion}\label{secDiscussion}
\subsection{Theories}
\begin{figure}[t]
  \begin{center}
  \includegraphics[width=6cm,height=5cm]{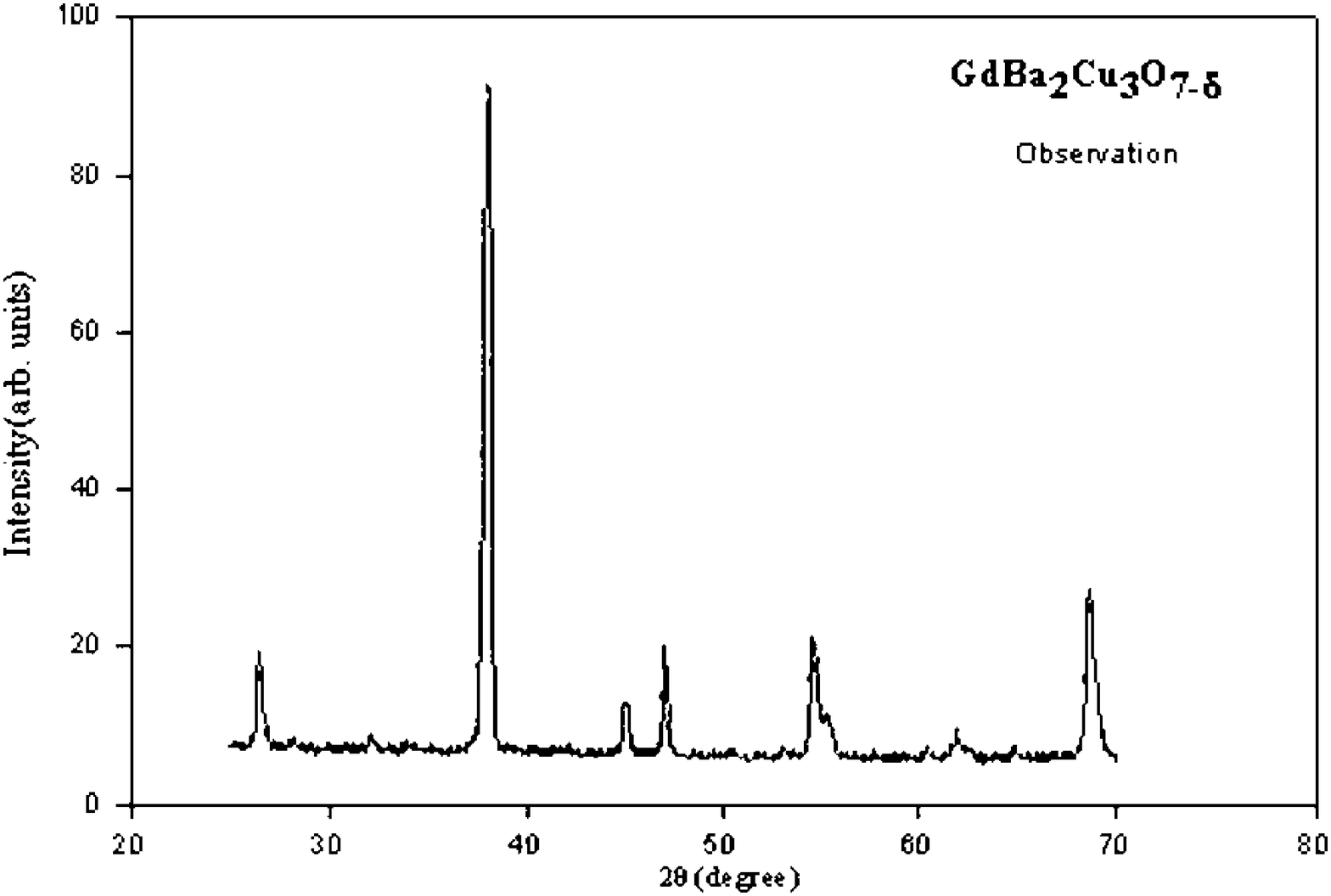}\hspace{0.5cm}
  \includegraphics[width=6cm,height=5cm]{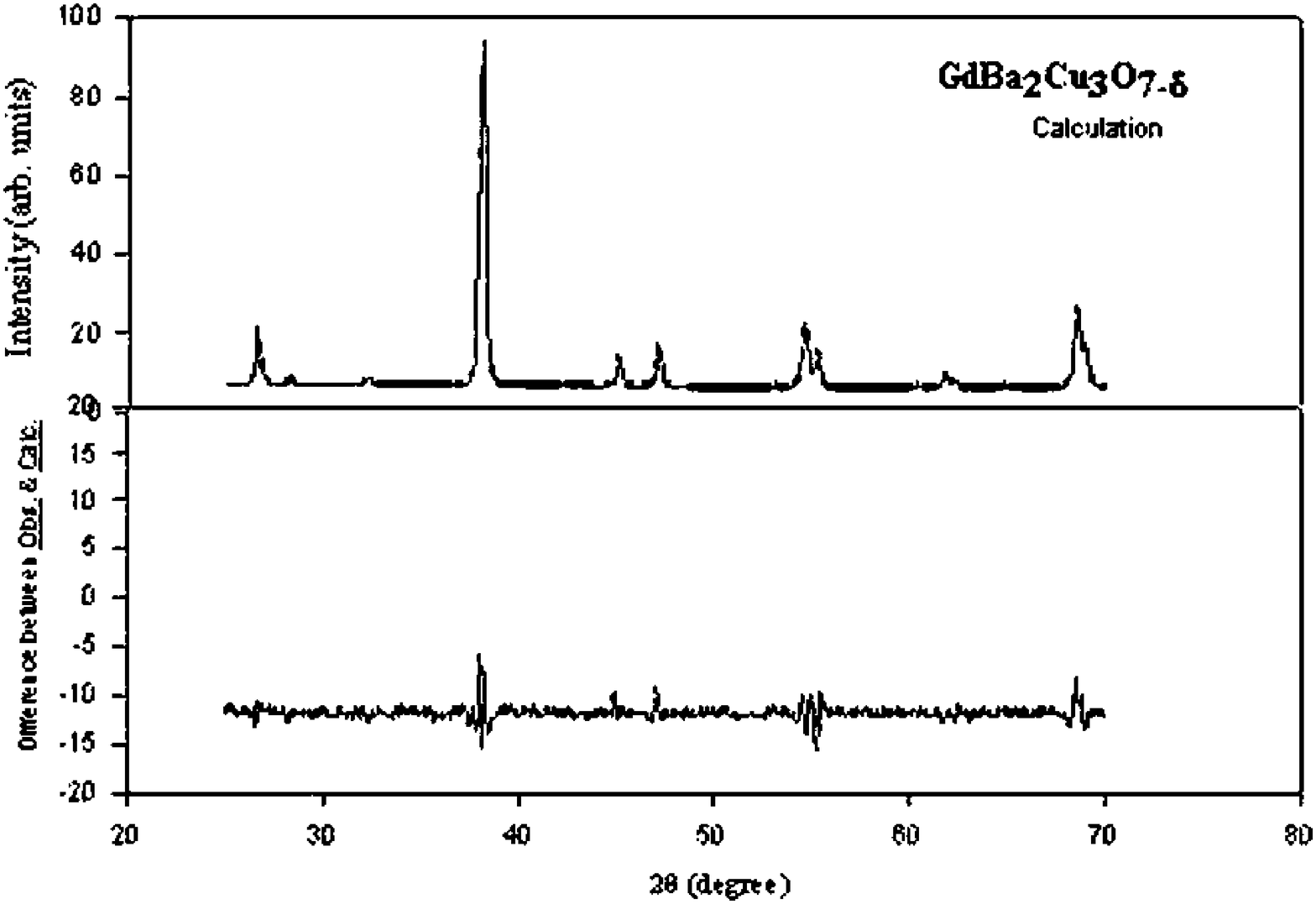}
  \end{center}
  \caption{XRD patterns; experimental (left), Ritveld analysis (upper right), Difference between experimental and Ritveld analysis(lower right).}
  \label{figRitveld}
\end{figure}
\begin{figure}[b]
  \begin{center}
  \includegraphics[width=6cm,height=5cm]{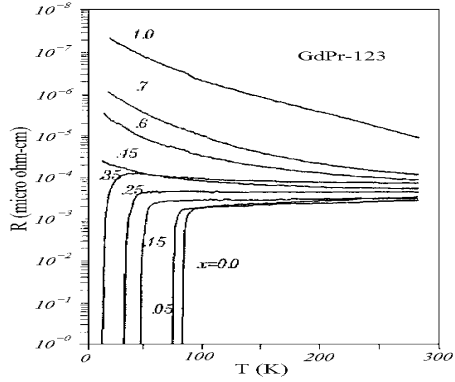}
  \end{center}
  \caption{The electrical resistivity of GdPr-123 versus temperature for different values of oxygen deficiency \cite{Akhavan3,Yamani2}.}
  \label{figRT}
\end{figure}
\begin{figure}
  \begin{center}
  \includegraphics*[width=6cm,height=5cm]{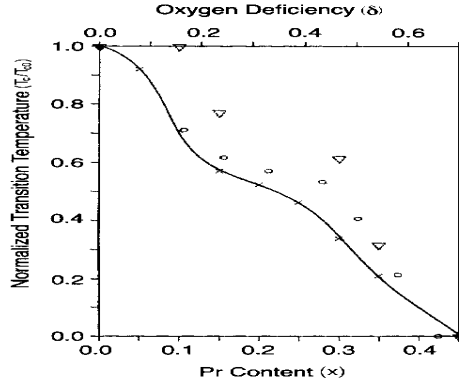}
  \end{center}
  \caption{The critical temperature of GdPr-123 versus Pr-doping value and oxygen deficiency $\delta$ \cite{Akhavan3}. Here, $T_{C0}$ is the transition temperature when $(x=0)$ and $(\delta=0)$.}
  \label{figTCx}
\end{figure}
\begin{figure}
  \begin{center}
  \includegraphics*[width=6cm,height=5cm]{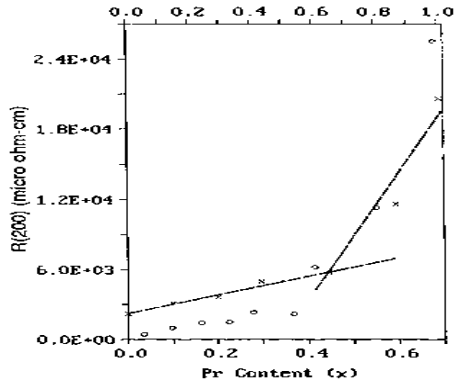}
  \end{center}
  \caption{The electrical resistivity of GdPr-123 versus Pr-doping value \cite{Akhavan3,Yamani2}.}
  \label{figRx}
\end{figure}
Orthorhombic structure phase of RPr-123 for Pr-doped value $(0<x<1)$
was constructed by standard solid state reaction method. Ritveld
analysis of XRD patterns for our samples shows in Fig.
\ref{figRitveld} that predominately phase of our samples is
orthorhombic. The lattice parameters, in average, are  $a = 3.84, b
= 3.89, c = 11.69 A^\circ$ Furthermore, it is shown that the volume
of unit cell increase by Pr-doping, without any change in phase of
the structure. This may be due to Pr
ion size which is greater than Gd ion size \cite{Akhavan3,Muroi,GCPD}.\\
The electrical resistivity of several samples have been shown in
Fig. \ref{figRT}. As is seen in this figure, by Pr-doping the
critical temperature, $T_C$, suppresses, and also the normal
resistivity increase. Figure. \ref{figTCx} shows suppression of
normalized transition temperature versus Pr-doping value $(x)$
\cite{Akhavan3}. It is emphasized that, there are two distinct
regime; normal and superconductivity, when Pr-doping value, $x$,
changes. Figure \ref{figRx} illustrates implicity these two regimes,
which are separated by the critical value $x_C=0.45$ for GdPr-123.\\
In addition, there are several anomalies appeared by Pr-doping, for
instance, increase of normal resistivity, no effect on Pinning
energy of vortices in vortex dynamics, and also no effect on
critical current density $J_C$ \cite{Akhavan3}.\\
There are some theories to explain Pr anomalies in 123-systems,
which also propose some explanations for $T_C$ suppression in
123-systems. There are listed in the following.
\subsubsection{Hole Filling}In this mechanism, it is considered that the valance of Pr is
significantly greater than +3. This, in turn, results to reduction
of hole concentration in the \CuO planes
\cite{Akhavan3,Matsuda,Gonacalves,Daadmehr,Muroi3}. However, the results
of Hall effect of measurements \cite{Matsuda}, magnetic
susceptibility \cite{Matsuda,Okai}, and some others
\cite{Akhavan3,Daadmehr} turned out inconsistent with some
spectroscopic data \cite{Radosky,Fink}.
\subsubsection{Pair Breaking}Strong exchange coupling of the Pr $4f$ moments with the spins of
the holes in \CuO planes results to magnetic pair breaking
\cite{Kebede,Neumeier,Maple}, It has been known that some rare-earth
elements such as Gd and Sm have greater $4f$ moments, however, they
do not result to suppression of superconductivity \cite{Akhavan3}.
\subsubsection{Mis-substitution}In 123-systems, Pr ions occupy Ba sites instead of Gd position
\cite{Blackstead}. However, by extended X-ray absorption fine
structure, it has been shown that Pr does not substitute for Ba in
Pr-123 Compounds in amounts greater than $1\%$ \cite{Booth,Harris}.
In addition, more recently, it has been observed that Pr/Ba
substitution in \GdBaPrd compounds does not affect its
superconductivity \cite{Mohammadizadeh3}.
\subsubsection{Hole Localization}Hole localization is based on strong hybridization between the Pr
localized $4f$ electron and the valance band states that are
associated with \CuO planes. The mobile hole on the O $2p_{\sigma}$
state tends to be in localized in O $2p_{\pi}$ state. Thus, the
localized hole cannot move, and gets immobile on \CuO planes. As a
result, the concentration of mobile holes on \CuO planes decreases.
On the other word, the hole localization occurs when two R sites
adjacent to a \CuO planes are occupied by Pr ions. Then Pr-doping
can decrease hole concentration on \CuO planes, and substantially,
$T_C$ is suppressed. Fehrenbacher and Rice (FR) \cite{Fehrenbacher}
by Hartree calculations, it has been shown that there are only two
stable electronic configurations in R-123 systems: i) all holes in a
\CuO plane are in the $\rm Cu^{+3}$ oxidation state, or exactly, in
hybridized Cu $3d$ - O $2p_\sigma$ orbital, and ii) all hole in a
\CuO plane are in the $\rm R^{+4}$ oxidation state, or exactly, in
hybridized R $4f$ - O $2p_\pi$ orbital. There is a small energy
difference between $\rm Pr^{3+}$ $4f$ level and the Fermi energy
$E_F$ which makes the hole transmission possible. This mechanism
provides a satisfactory explanation for Pr effects and insulating
nature of \RPrd \cite{Muroi2,Maple}.
\subsubsection{Hole Clustering}Here, we show that an alternative mechanism, which is a modification
of the hole localization, is also possible. In each superconducting
cell, hole concentration decreases when Pr ions substitute for Y(Gd)
\cite{Akhavan3,Muroi}. By increasing Pr-doping value, the
probability of that the insulator cells are near some others
increases. In this way, some insulator clusters may be formed. The
holes trapped inside such these clusters become immobile, and
subsequently the hole concentration effectively reduces. Moreover,
we believe that when a (nearly) closed belt of such insulator cells
is formed, the holes trapped inside get immobile and effectively
reduce the hole concentration. This event is similar to the case in
which all inside cells have Pr impurities.\\
In the following sub section, we provide a geometrical background,
based upon band percolation theory, for modeling hole localization
and hole clustering effects.

\subsection{Band Percolation modeling in 123-systems}In this part, we consider some assumptions on the base of which we
establish a 2D band percolation model for random distribution of Pr
within \CuO superconductive planes \cite{Stauffer,Hoshen}. The main
reason for dimensionally 2 for percolation model is that 123-systems
are highly anisotropic so that the anisotropy parameter (which is
relate to the effective mass tensor) ,$\gamma$, ranges from 7 to 50
\cite{Tinkhham}. Our assumptions are as follow:\\
\begin{enumerate}
\item In the absence of Pr, to have superconductivity we need at least a CuO2 cell with local hole concentration of $0.25$
\cite{Takenaka}.\\
\item An oxygen ion acts as a hole trap in a \CuO
plane when its two neighboring R sites are both occupied by Pr
ions.\\
\item Such oxygen ions, irrespective of that they are
trapping holes, are inaccessible to mobile holes. In this way, they
can be regarded as point defects \cite{Muroi2}.\\
\item Two basic units merge to form a single
superconducting region when they are adjacent to each other, sharing
two Cu ions and one oxygen ion when they are separated by a
non-superconducting layer.\\
\item Tendency of R and Pr ions to make a separate
structural phase in \CuO planes increases by increasing R ion size,
so that large ions tend to make more compact larger clusters
\cite{Muroi,GCPD,Malik}. So only for some rare-earth elements like
Gd or Y the distribution of Pr position in superconducting \CuO
planes is uniform random \cite{Booth}.\\
\end{enumerate}
Validity of the above mentioned assumptions have been discussed in
\cite{Muroi2} and the references there in. Therefore, \CuO planes
are made up of two distinct parts: i) S(superconducting)-cells:
these are defect free \CuO cells which have been considered by the
assumption 1 as superconducting. According to the assumption 2, when
such these cells get adjacent to each other a superconducting
cluster is formed. ii) N(normal)-cells: these are cells in which Pr
ions substitute for Gd or Y ions. Also, by combination of N-cells a
N-cluster is formed.\\
By the assumptions 1-4, it is concluded that the probability for
that an oxygen ion on a \CuO plane have two Pr ions in its
neighbored (in plane of R ions) is $x^2$. By considering that there
are 4 nearest neighbors for any site, and provided the oxygen
entraps one hole, mobile hole concentration in \CuO planes is as
follows.
\begin{figure}[t]
  \begin{center}
    \includegraphics*[scale=0.6]{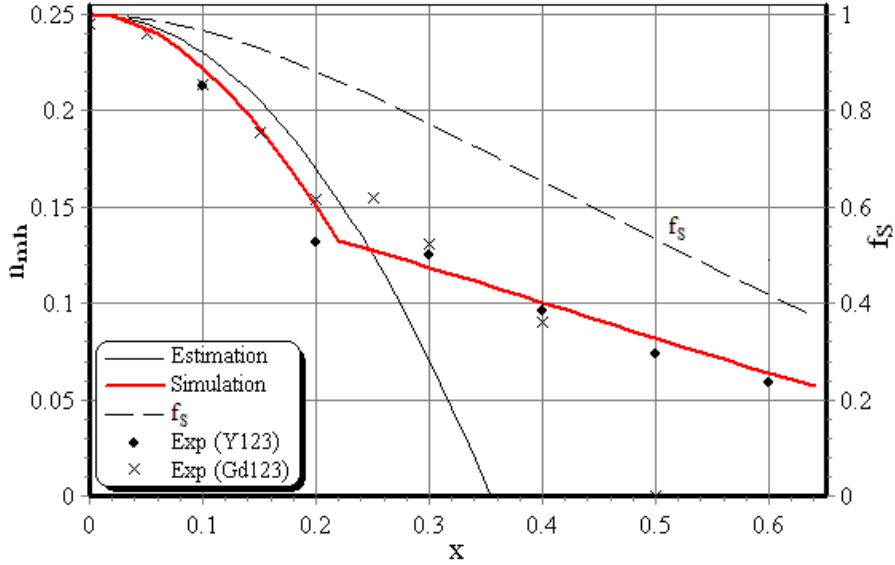}
  \end{center}
  \caption{Mean field approximation hole localization obtained form Eq. \ref{eqnmh} (thin solid curve), the simulation result (thick solid curve), and experimental result
for Gd-123 and Y-123 systems \cite{Mohammadizadeh,Lee}. The dashed
curve (measured by the right axis) shows fraction of S-clusters
$(f_S)$.} \label{figPercol1}
\end{figure}
\begin{figure}[b]
  \begin{center}
    \includegraphics*[scale=0.6]{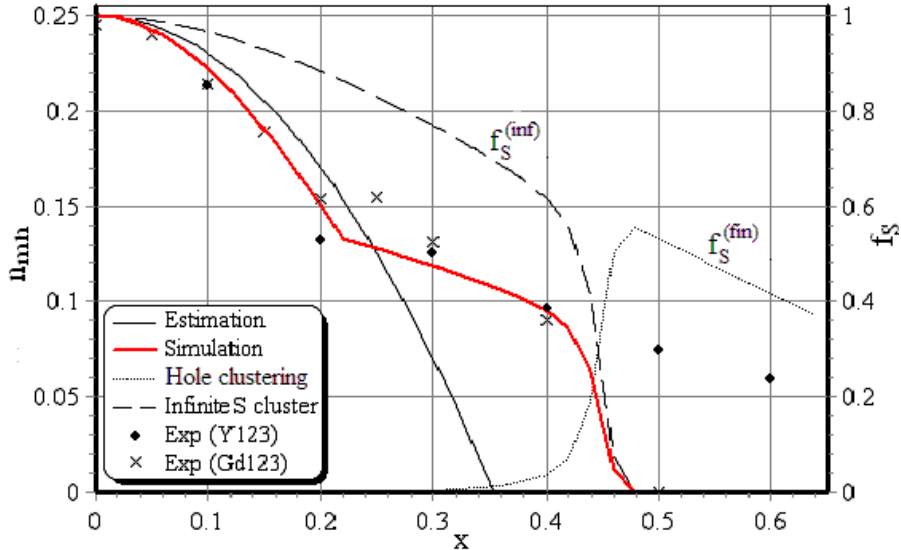}
  \end{center}
  \caption{Hole concentration vs. Pr-doping value, obtained by
including hole clustering effect. In $x \approx 0.45$, hole
concentration experiences a falloff (thick solid curve). Fraction of
infinite S-clusters is shown by the dotted line, and is measured by
the right axis.} \label{figPercol2}
\end{figure}
\begin{equation}\label{eqnmh}
  n_{mh} = 0.25 - 2x^2
\end{equation}
This relation presents a good approximation for hole concentration
on \CuO plane for lower doping values, $x$, however, for other
values there is not such agreement \cite{Muroi2}. The experimental
results for Y-123 and Gd-123 systems have been shown in Fig.
\ref{figPercol1}. However, a point, here, is worth noting. In Fig.
\ref{figPercol1}, apparently the experimental data show a less hole
concentration. Indeed, as we have explained in Appendix, this effect
is due to existence of N-clusters in the system. If we make a
simulation and calculate the $n_{mh}$ only on the basis of above
Assumptions, we will find the mobile hole concentration only
$n_{mh}(x) > 0.15$ with a good approximation (Fig.
\ref{figPercol1}). The Coulomb repulsion between the localized holes
tries to prevent localization of holes after the value of Pr-doping
approaches $n_{mh}(x) \thickapprox 0.15$. The effects of hole
localization descend after this point. In this stage, the size of
normal clusters increases by increasing $x$. The two phase behavior
illustrated in Fig. \ref{figPercol1}, is similar to the phase
transition from ortho. 1 to ortho. 2 observed in behavior of $T_C$
in terms of oxygen deficiency in 123-systems \cite{Daadmehr}. When
Pr-doping value is small, it is natural that the size of N-clusters
are small so that effectively no S-cluster can be present inside
N-clusters. By increasing $x$, the size of N-clusters grows so that
it is likely now that some S-clusters get trapped inside N-clusters.
In this case, the effective value of hole concentration decreases.
Another case, also, may happen. It is  possible that the formed belt
of N-clusters is not closed. However, when the separation of the two
ends of such a belt is small in our investigation, by a good
approximation, it can be considered closed. In other words, we
assume that a nearly closed belt of insulator cells is taken to be
closed. Therefore, such these belts exclude all the holes within.
Both of the cases cause some mobile holes get caught inside a region
bounded with N-clusters, and thus the contribution of mobile holes
in conductivity decreases. When Pr-doping tends to the critical
value $(x_C)$ this effect takes place very fast, and we expect to
see a phase transition near the percolation threshold (Fig.
\ref{figPercol2}). In this region, as is evident from Fig.
\ref{figPercol2}, as small change of $x$ value gives rise to a
considerable change in amount of mobile hole concentration. Decrease
of the mobile hole concentration can be attributed to the decrease
of the size of infinite S-clusters in our percolation model. Figure
\ref{figPercol2} contains the experimental result for Y-123 and
Gd-123 systems.

\section{Conclusion}\label{secConclusion}In this work, we want to explain $T_C$ suppression in 123-systems
based on a geometrical modeling. To this aim, some HTSC samples of
Gd-123 with dominant orthorhombic phase, for different doping values
$0<x<1$, have been prepared by solid state reaction technique. Then
we have characterized the samples by XRD and Ritveld analysis. By
making some assumption supported with band percolation theory, we
have investigated suppression of superconductivity and behavior of
hole concentration on \CuO planes in relation of pr-doping. It has
been shown that in low $x$ values, the obtained hole concentration,
after an important correction, show a good agreement with
experiment. By increasing the doping value, the effect of Coulomb
repulsion between trapped holes become more important so that it can
compete with hole localization effect, and subsequently, limits it.
Finally, it has been obtained that in high values of Pr-doping hole
clustering is the dominant effect in explanation of special behavior
of $n_{mh}$, $T_C$ suppression, and phase transition versus $x$.



\appendix
\section{Appendix}
\begin{figure}[t]
  \begin{center}
  \includegraphics*[width=6cm,height=4cm]{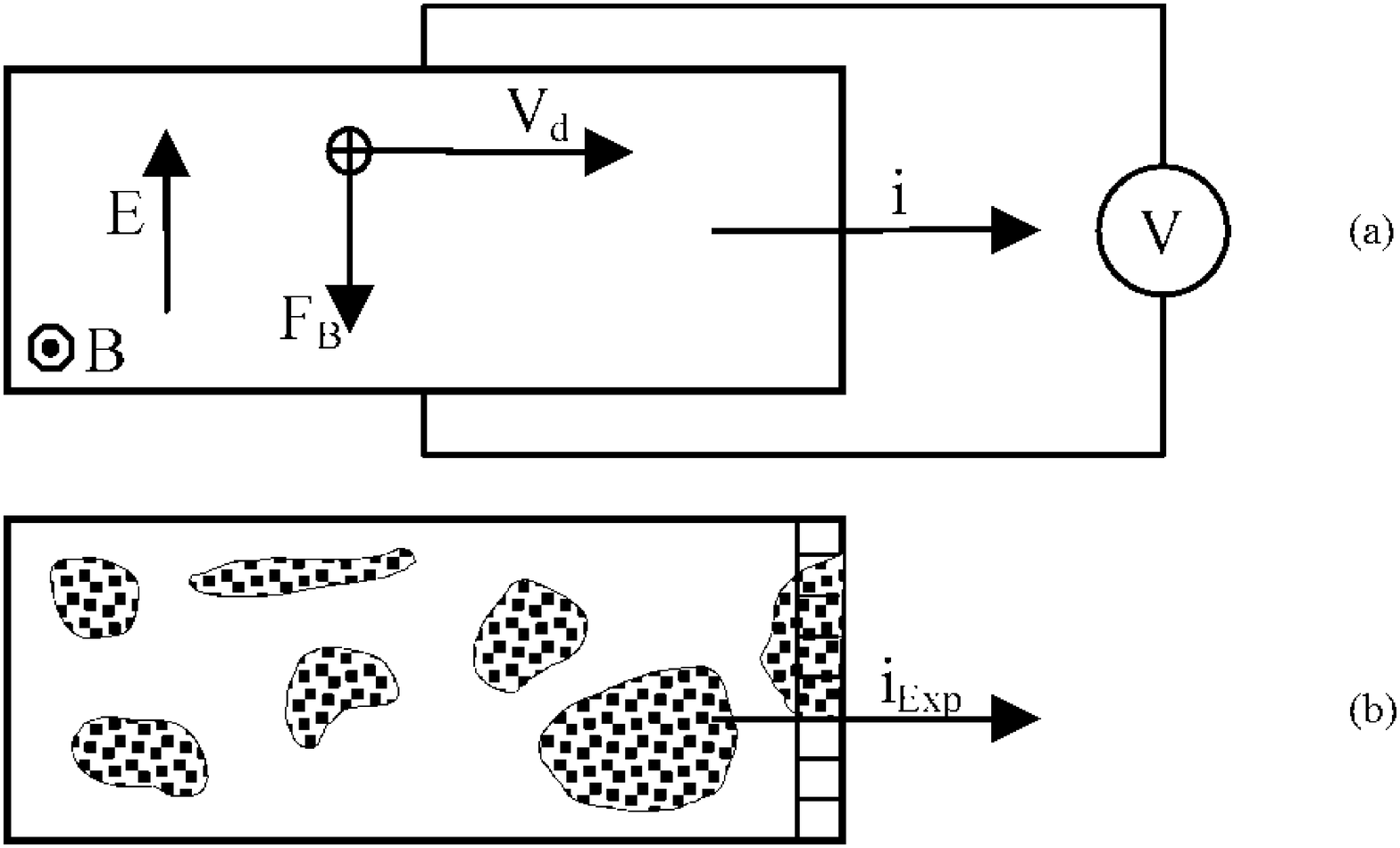}
  \end{center}
  \caption{"(a) Hall effect setup for a pure metal. (b) Surface of N-clusters, in superconductivity planes, is avoided by current."}\label{figHall}
\end{figure}

To compare the experimental result of hole concentration based on
Hall effect with the accepted definition for $n^{(S)}_{mh}$ of
S-clusters in \CuO planes \cite{Muroi3}, defined as
\begin{equation}\label{eqn}
  n^{(S)}_{mh}=\frac{N_{mh}}{V_S}=\frac{N_{mh}}{V.f_S}
\end{equation}
one must take the measurement mechanism of hole concentration into
account. I fact, in Hall effect measurements to find carrier concentration it is
implicity assumed that the sample is a pure metal (Fig.
\ref{figHall}). Thus, the following relation for carrier
concentration is obtained:
\begin{equation}\label{eqHall}
  n=\frac{Bi}{etV}
\end{equation}
However, in our experiments the processed samples are granular. By Atomic Force Microscopy (AFM)
imaging, or other relevant techniques, and using metal-insulator-metal (M-I-M) models the
parameters for no-granular single crystal are obtained \cite{Toker}.
Yet, even after this correction, the effect of existence of N-cluster are
ignored. This simply means that the actual current measured in
experiment is less than the value obtained when the whole compound
pure isotropic metal. Thus, to enter this effect into play (Fig.
\ref{figHall}), the current $i$ should be considered as
\begin{equation}\label{eqiExp}
  i_{Exp}=i.f_S
\end{equation}
equation (\ref{eqHall}) and (\ref{eqiExp}), now, give
\begin{equation}\label{eqnmh2}
  n=\frac{Bi_{Exp}/f_S}{etV}
\end{equation}
This is, the fraction of S-clusters, $f_S$, should be multiplied by
$n_{mh}$ to give the correct experimental value:
\begin{equation}\label{eqExpnmh}
  n_{mh}=n_{mh}^{(Exp)}/f_S
\end{equation}


\end{document}